\documentclass[a4paper]{article}
\usepackage{color,graphicx}
\usepackage{amsmath}
\usepackage{amsthm}
\usepackage{amssymb}
\usepackage{amsfonts}
\RequirePackage{ifpdf} 
\ifpdf
\usepackage[pdftex,bookmarks=true,
		   bookmarksnumbered=false,
		   bookmarksopen=false,
		   colorlinks=true,
		   linkcolor=webred] {hyperref}
\definecolor{webgreen}{rgb}{0, 0.5, 0} 
\definecolor{webblue}{rgb}{0, 0, 0.5} 
\definecolor{webred}{rgb}{0.5, 0, 0}   
\else
\newcommand{\href}[2]{ #1 }
\fi
\newcommand{\T}[1]{\textrm{#1}}
\title{Quantum Shortest Path Netsukuku}
\author{AlpT (@freaknet.org)}
\author{http://netsukuku.freaknet.org\\AlpT (@freaknet.org)}
\begin{document}
\maketitle

\begin{abstract}
	This document describes the QSPN, the routing discovery algorithm used
	by Netsukuku.
	Through a deductive analysis the main proprieties of the QSPN are
	shown. Moreover, a second version of the algorithm, is presented.
\end{abstract}
\pagenumbering{roman}
\pagebreak
\begin{small}
  This document is part of Netsukuku.\\
  Copyright \copyright 2007 Andrea Lo Pumo aka AlpT $<$alpt@freaknet.org$>$.
  All rights reserved.

  This document is free; you can redistribute it and/or modify it
  under the terms of the GNU General Public License as published by
  the Free Software Foundation; either version 2 of the License, or
  (at your option) any later version.

  This document is distributed in the hope that it will be useful, but
  WITHOUT ANY WARRANTY; without even the implied warranty of
  MERCHANTABILITY or FITNESS FOR A PARTICULAR PURPOSE\@.  See the GNU
  General Public License for more details.

  You should have received a copy of the GNU General Public License
  along with this document; if not, write to the Free Software
  Foundation, Inc., 675 Mass Ave, Cambridge, MA 02139, USA.
\end{small}

\clearpage
\tableofcontents
\clearpage
\pagenumbering{arabic}

\section{Preface}
\label{sec:preface}

The first part of the document describes the reasoning which led us to the
construction of the current form of the QSPN v2.
If you are just interested in the description of the QSPN v1 and v2 and you
already know the concept of the Tracer Packet, you can directly skip to
section \ref{sec:CTP}.

\section{The general idea}
\label{sec:general_idea}

The aim of Netsukuku is to be a (physical) scalable mesh network, completely
distributed and decentralised, anonymous and autonomous.

The software, which must be executed by every node of the net, has to be
unobtrusive. It has to use very few CPU and memory resources, in this way it
will be possible to run it inside low-performance computers, like Access Points,
embedded devices and old computers.

If this requirements are met, Netsukuku can be easily used to build a worldwide
distributed, anonymous and not controlled network, separated from the
Internet, without the support of any servers, ISPs or control authorities.

\subsection{The network model}
\label{sec:net_model}

Netsukuku prioritises the stability and the scalability of net: the network
has to be able to grow to even $2^{2^7}$ nodes.

A completely dynamic network would requires rapid and frequent updates
of the routes and this is in contrast with the stability and the scalability
requirements of Netsukuku.
For this reason, we restrict Netsukuku to the case where a node won't change
its physical location quickly or often.

This assumption is licit, because the location of a wifi node mounted on
top of a building won't change and its only dynamic actions would be the
joining and the disconnection to and from the network and the changes of the
quality of its wifi links.
However, there are some consequences of this assumption:
\begin{enumerate}
	\item	Mobiles node aren't supported by Netsukuku algorithms.
		\footnote{It is possible to use other mesh network protocols
		designed for mobility in conjunction with Netsukuku, in the
		same way they are used in conjunction
		with the Internet (f.e. see \href{http://olsrd.org}{olsrd}). }
	\item   The network isn't updated quickly: several minutes may be
		required before all the nodes become aware of a change of the
		network (new nodes have joined, more efficient routes have
		become available, \dots). However, when a node joins
		the network, it can reach all the other nodes from the first
		instant, using the routes of its neighbours.
\end{enumerate}

\subsection{The routing algorithm}
One of the most important parts of Netsukuku, is the routing discovery
algorithm, which is responsible to find all the most efficient routes of the
network. These routes will permit to each node to reach any other node.

The routing algorithm must be capable to find the routes without overloading
the network or the nodes' CPU and memory resources.

\subsection{The QSPN}

Netsukuku implements its own algorithm, the \emph{QSPN} (\textbf{Q}uantum
\textbf{S}hortest \textbf{P}ath \textbf{N}etsukuku). The name derives from the
way of working of its principal component: the \emph{TP} (Tracer Packet), a
packet which gains a ``quantum'' of information at each hop.

The QSPN is based on the assumptions described in section \ref{sec:net_model}.

\section{Network topology}
\label{sec:net_topology}

The QSPN alone wouldn't be capable of handling the whole network, because it
would still require too much memory. For example, even if we store just one
route to reach one node and even if this route costs one byte, we would need
1Gb of memory for a network composed by $10^9$ nodes (the current Internet).

For this reason, it's necessary to structure the network in a convenient
topology.

\subsection{Fractal topology}
\label{sec:fractal_topology}
Netsukuku, adopts a fractal like structure:
256 nodes are grouped inside a \emph{group node} (gnode), 256 group nodes are grouped
in a single \emph{group of group nodes} (ggnode), 256 group of group nodes are
grouped in a gggnode, and so on.
(We won't analyse the topology of Netsukuku. You can find more information
about it in the proper document: \cite{ntktopology}).
\newline
Since each gnode acts as a single real node,
the QSPN is able to operate independently on each level of the fractal.

Since in each level there are a maximum of 256 (g)nodes, the QSPN will
always operate on a maximum of 256 (g)nodes, therefore we would need just to
be sure that it works as expected on every cases of a graph composed by $\le
256$ nodes. By the way, we'll directly analyse the general case.

For the sake of simplicity, in this paper, we will assume to operate on level
0 (the level formed by 256 single nodes).

\section{Tracer Packet}
\label{sec:TP}

A \emph{TP} (Tracer Packet) is the fundamental concept on which the QSPN is
based: 
it is a packet which stores in its body the IDs of the traversed hops.

\subsection{Tracer Packet flood}
\label{sec:TP_flood}

A TP isn't sent to a specific destination but instead, it is used to flood the
network. By saying ``the node A sends a TP'' we mean that ``the node A is
starting a TP flood''.

A TP flood passes only once through each node of the net: a node which
receives a TP will forward it to all its neighbours, except the one from which
it received the TP. Once a node has forwarded a TP, it will not forward any
other TPs of the same flood.

\subsection{Proprieties of the tracer packet}
\label{sec:proprieties_TP}

\begin{enumerate}
	\item A node $D$ which received a TP, can know the exact route covered
		by the TP. Therefore, $D$ can know the route to reach the
		source node $S$, which sent the TP, and the routes to reach
		the nodes standing in the middle of the route.
		
		For example, suppose that the TP received by $D$ is: $\left\{
		S, A, B, C, D \right\}$. By looking at the packet $D$ will
		know that the route to reach $B$ is $C\rightarrow B$, to reach $A$ is
		$C\rightarrow B\rightarrow A$, and finally to reach $S$ is
		$C\rightarrow B\rightarrow A\rightarrow S$.
		The same also applies for all the other nodes which received
		the TP, f.e, $B$ knows that its route to reach $S$ is
		$A\rightarrow S$.
	\item The \emph{bouquet of $S$} is the set of all the TPs which will
		be forwarded or sent by the node $S$ during the flood.
		The first TP of this bouquet received by a generic node $D$,
		will be the TP which covered the fastest route which connects
		$S$ to $D$.
		The fastest $S \rightarrow D$ route is the route with the
		minimum \emph{rtt} (Round-Trip Time) between $S$ and $D$.
		This property is also valid if $S$ is the node which started
		the TP flood, i.e. the first node which sent the first bouquet
		of the TP flood.
\end{enumerate}

\subsubsection*{Example}
\begin{figure}[h]
	\begin{center}
		\includegraphics[scale=0.4]{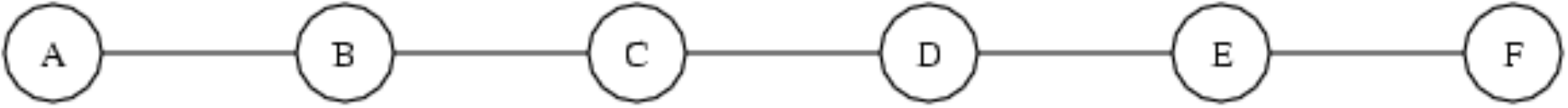}
	\end{center}
	\caption{A simple graph}
\end{figure}

Suppose that $D$ sends a TP. The TP will cover this routes:
$D \rightarrow E \rightarrow F$ and $D \rightarrow C \rightarrow B \rightarrow A$.
When the TP reaches the node $F$ and the node $A$, the flood will stop,
because either $A$ and $F$ won't be able to forward the TP to any other node.

At the end, $A$ will know the route $A \rightarrow B \rightarrow C \rightarrow D$ and $F$ will know the
route $F \rightarrow E \rightarrow D$.

\section{Routes of a graph}
\label{sec:gen_routes}

Given a graph $\mathbf{G}$ we want to find all the existing routes between a node and
all the other nodes.

Let $N$ be a generic node. Starting from $N$ we explore the entire graph
until we re-enter in a cycle already visited or we cannot proceed any further.
This approach is similar to the Depth-First Search\cite{DFS} algorithm, but instead of
searching for a specific goal, we just traverse the entire graph.
Note that a cycle is traversed only once, because we need non redundant
routes. In other words, if we already know the $S \rightarrow A \rightarrow B
\rightarrow C \rightarrow D$ route,
it's useless to known that we can reach $D$ with the $S \rightarrow A
\rightarrow B \rightarrow C \rightarrow A \rightarrow B \rightarrow C
\rightarrow D$ route.

This is the pseudo code of the algorithm:

\begin{verbatim}
generate_routes(G) {
        forall node in G
                /* Starts the exploration of the graph from the ``node'' of the
                   graph ``G'' and print all its routes */
                walk(node, node)
}

/* Print all the routes which start from the node `N' */
walk(N, branch) {
        deepened=0

        forall L in N.links
                /* L is a neighbour of N */

                if(L in branch)
                        /* If ``L'' is already contained in the explored
                           branch, we've found a cycle. Since we just need to
                           traverse only once a cycle, we skip this ``L'' node
                           and continue to consider the other neighbours
                           of N */
                        continue;

                newbranch=branch + L    /* Append in the explored branch the
                                           ``L'' node. */

                walk(L, newbranch)      /* Recursively  explore the new
                                           branch */
                
                /* Indicate that we've deepened in the graph at least once */
                deepened=1

        if(!deepened)
                /* We haven't deepened in the above for, this means that the
                   current branch can't be explored anymore, therefore it is a
                   valid route. Print it */
                print branch
}
\end{verbatim}

A proof of concept of the above algorithm has been implemented in Awk
\cite{genrouteawk}.

\subsection*{Example}
Consider this graph:

\begin{figure}[h]
	\begin{center}
		\includegraphics[scale=0.4]{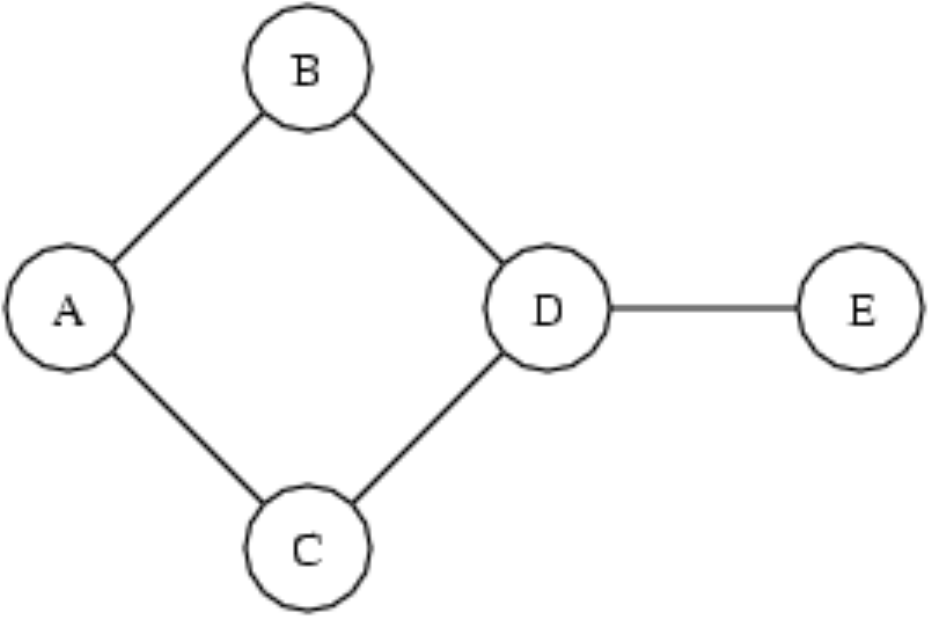}
	\end{center}
	\caption{A simple graph with one segment and one cycle}
	\label{fig:gen_route_sample}
\end{figure}

Given this graph as input the algorithm will output:
\label{sec:genroute_output}
\begin{align*}
& A \rightarrow B \rightarrow D \rightarrow C\\
& A \rightarrow B \rightarrow D \rightarrow E\\
& A \rightarrow C \rightarrow D \rightarrow B\\
& A \rightarrow C \rightarrow D \rightarrow E\\
& B \rightarrow A \rightarrow C \rightarrow D \rightarrow E\\
& B \rightarrow D \rightarrow C \rightarrow A\\
& B \rightarrow D \rightarrow E\\
& C \rightarrow A \rightarrow B \rightarrow D \rightarrow E\\
& C \rightarrow D \rightarrow B \rightarrow A\\
& C \rightarrow D \rightarrow E\\
& D \rightarrow B \rightarrow A \rightarrow C\\
& D \rightarrow C \rightarrow A \rightarrow B\\
& D \rightarrow E
\end{align*}

\section{Acyclic Tracer Packet flood}
\label{sec:acyclic_TP_flood}

We can consider each route given by the output of the above algorithm as a
single Tracer Packet.
In fact, it is possible to implement the same algorithm using a slightly
modified version of the TP flood, called the Acyclic TP flood:

The flood is not restricted like in a normal TP flood: one or more ATP can
pass from the same node. The end of the flood is given by this rule: a node
will not forward to any of its neighbours the ATP if its node ID is already present
in the route contained in the body of the packet. With this rule an ATP can
walk in a cycle only once, hence the name.
Finally, like in the normal TP, a node doesn't forward the ATP to the
neighbour from which it has received the packet itself.

If every node of the network sends an ATP flood, then every node will get
all the possible routes to reach any other node.
\newline
As you can see, the ATP flood performs a ``live'' version of the algorithm
described in section \ref{sec:gen_routes}.
Obviously this is far from an efficient routing discovery algorithm, but it
represents a good start.

\section{Routes simplification}
\label{sec:simplify_routes}

Looking carefully at the example output (\ref{sec:genroute_output}) of the
Generate Route algorithm, we can notice that many routes are higly redundant,
in other words, some routes are almost the same.
Consider for example the following four routes:
\begin{align}
	& A \rightarrow B \rightarrow D \rightarrow E \label{eq:e1}\\
	& D \rightarrow E \label{eq:e2} \\
	& A \rightarrow B \rightarrow D \rightarrow C \label{eq:e3}\\
	& D \rightarrow C \rightarrow A \rightarrow B \label{eq:e4}
\end{align}

As we've seen in the previous section \ref{sec:acyclic_TP_flood}, we can consider
these routes as effective Tracer Packets.
In this example, the TP (1) cover the same route of the TP
(2). Therefore we can save one TP by just sending the TP
(1), which will traverse the route (2) too.

The TP (3) covers part of the TP (4), thus we can simplify
the two of them by just sending a TP which cover this route: $A \rightarrow B
\rightarrow D \rightarrow C \rightarrow A \rightarrow B$.

Continuing in this process we can further simplify the two TP:
\[ ABDCAB + ABDE \Rightarrow  ABDCABDE \]

Thus, from the initial four TPs we've found a unique TP which gives the same
routes of the original ones.

\subsection{Simplification rules}
\label{sec:simplification_rules}

We can derive some rules to simplify routes.

Since we can represent a route as a string where each symbol is a node, we can
also describe the routes simplification as a series of operations on strings.

In the following rules, each letter found in an expression represents a generic string,
which may be also the NULL string, f.e. the ``$XX$'' string can be anything like
$foofoo$ or $1234512345$.
\\
The $c\dots c$ expression represents a cycle, where the $c$ character refers
to just one node, and not to an entire string.
\begin{description}
	\item[XY+YZ $\Rightarrow$ XYZ]
		If two routes share respectively the ending and the starting
		part, they can be merged into a unique route. Example:
		\[ABCDE + CDEKRE \Rightarrow ABCDEKRE \]
	\item[YXZ + X $\Rightarrow$ YXZ]
		Example:
		\[123ABCXYZ + ABC \Rightarrow 123ABCXYZ\]
	\item[Xc\dots c + XcY $\Rightarrow$ Xc\dots cY]
		Example:
		\[123ABCDA + 123A987 \Rightarrow 123ABCDA987\]
	\item[c\dots cZ + YcZ $\Rightarrow$ Yc\dots cZ]
		Example:
		\[ABCDA123 + 987A123 \Rightarrow 987ABCDA123\]
	\item[c\dots c + YcZ $\Rightarrow$ Yc\dots cZ]
		Example:
		\[ABCDA + 987A123 \Rightarrow 987ABCDA123\]
	\item[Invalid route]
		\label{sec:simroute_invalid}
		A route must not be in the form of:
		\[ XacaY \]
		where $a$ and $c$ are two nodes.
		A simplification, which gives a route of this form, is
		not considered valid. This is because a TP must not change its
		verse while traversing a network.
\end{description}

All these rules can be applied recursively to the routes of a graph, until
they cannot be simplified anymore.

A proof of concept of the above algorithm has been implemented in Awk \cite{simrouteawk}.

\subsection*{Example}
Simplifying all the routes of the example \ref{sec:genroute_output}, we obtain
just these two TPs:
\begin{align}
 &A \rightarrow B \rightarrow D \rightarrow C \rightarrow A \rightarrow B
 \rightarrow D \rightarrow E\\
 &A \rightarrow C \rightarrow D \rightarrow B \rightarrow A \rightarrow C \rightarrow D \rightarrow E
\end{align}

You can verify that all the routes listed in \ref{sec:genroute_output} are
contained in these two simplified TPs.

\subsection{General results}
By looking at many different simplifications, we can recognize some general
rules:
\begin{enumerate}
	\item For each TP there has to be its inverse. For example, if there's
		a TP which covered the route $12345$, then there has to be at
		least the TP which covers the inverse route $54321$.
	\item In a segment, to give all the routes to all the nodes, it is
		sufficient that the two extremes sends a TP. Example:
		\begin{figure}[h]
			\begin{center}
				\includegraphics[scale=0.4]{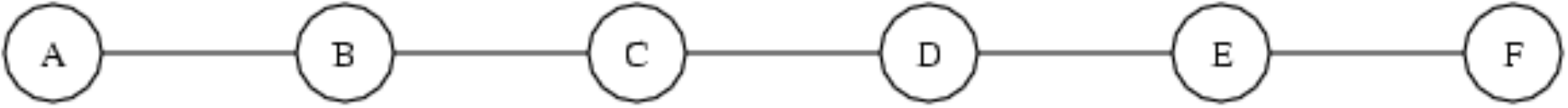}
			\end{center}
			\caption{A segment}
		\end{figure}
		in this case, if $A$ and $F$ send a TP, all the routes will be
		generated, since the two TP would be: $ABCDEF$ and $FEDCBA$.
		You can verify that in these two TP, there are contained all
		the routes of the segment.
	\item In a cycle, just two TP are needed, and one is the reverse of
		the other. The first can be constructed in this way: 
		\begin{itemize}
			\item Choose a node of the cycle, this will be the
				pivot node.
			\item Start from one neighbour of the pivot and write
				sequencially all the other nodes until you
				return to the pivot (but do not include it).
				Call this string $C$.
			\item The TP will be:
				\[CpC\]
				where $p$ is the pivot node.
		\end{itemize}
		Example:
		\begin{figure}[h]
			\begin{center}
				\includegraphics[scale=0.4]{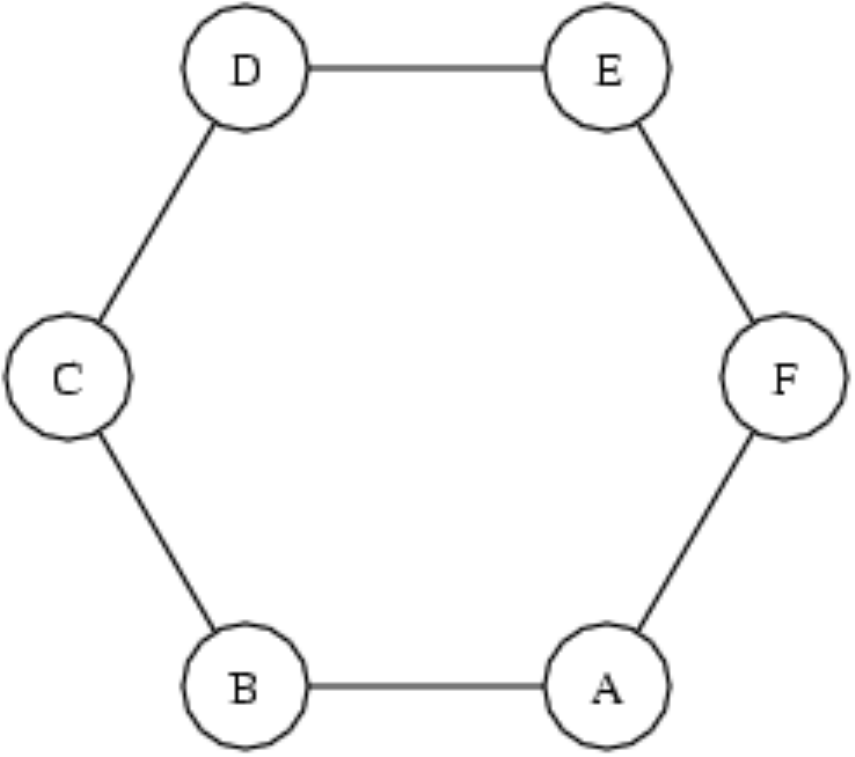}
			\end{center}
			\caption{A cycle}
		\end{figure}
		if we choose the node $D$ as the pivot, we can write the TP
		as:
		\[ EFABCDEFABC \]
		and its reverse:
		\[ CBAFEDCBAFE \]
		These two TPs will give all the routes to all the nodes of the
		cycle.
\end{enumerate}

\subsection{The question}

Can we implement a ``live'' version of the Simplify Route algorithm like we
did with the Generate Route one?\\
The reply is ahead.

\newpage
\section{Continuous Tracer Packet}
\label{sec:CTP}

A Continuous Tracer Packet (CTP) is an extension of the TP flood: a node will
always forward a TP to all its neighbours, excepting the one from which it has
received the TP.
If a node is an extreme of a segment, i.e. a node with just one link, it will
erase the route stored in the body of the TP and will forward back the TP.

In short, a CTP is a TP flood which will never end, thus it will continue to
explore all the infinite combination of routes. 

\subsection*{Example}
Consider this graph.
\begin{figure}[h]
	\begin{center}
		\includegraphics[scale=0.4]{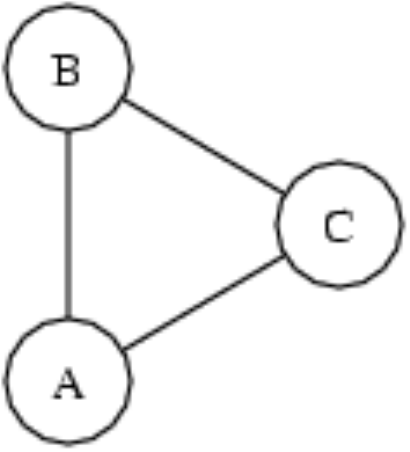}
	\end{center}
	\caption{The simplest cyclic graph}
	\label{fig:cycleABC}
\end{figure}
If $A$ sends a CTP flood, there will be two CTPs that will explore
respectively these routes:
\[
	 A \rightarrow B \rightarrow C \rightarrow A \rightarrow B \rightarrow
	 C \rightarrow A \rightarrow B \rightarrow C \rightarrow A \rightarrow
	 B \rightarrow C \rightarrow \dots
\]
\[
	 A \rightarrow C \rightarrow B \rightarrow A \rightarrow C \rightarrow
	 B \rightarrow A \rightarrow C \rightarrow B \rightarrow A \rightarrow
	 C \rightarrow B \rightarrow \dots
\]

\subsection{Reflected CTP}
\label{sec:reflected_CTP}

Suppose that the node $N$ has just one link.
$N$, before back forwarding the received CTP, erases the route contained in
the body, because the nodes preceding it, already know this same route.
\\
For example, consider this segment:
\[ \cdots \leftrightarrow A \leftrightarrow B \leftrightarrow C \leftrightarrow N \]
If $N$ hadn't erased the route received in the CTP, $A$ would have received
the following CTP:
\[
\cdots \rightarrow A \rightarrow B \rightarrow C \rightarrow N \rightarrow C \rightarrow B \rightarrow A
\]
This packet contains the route $C \rightarrow N \rightarrow C$, which is
invalid, as explained in section \ref{sec:simroute_invalid}.
The valid parts of the packet are: $\cdots \rightarrow A \rightarrow B
\rightarrow C \rightarrow N$ and $N \rightarrow C \rightarrow B \rightarrow
A$. For this reason, when $N$ receives the first part, it will send a new,
empty CTP.

\section{QSPN v2}
\label{sec:QSPNv2}
The second version of the QSPN\footnote{The short name of the QSPN v2 is
$Q^2$} can be described in a single phrase:

\emph{A Continuous Tracer Packet will continue to roam inside}

\emph{the network until it carries interesting information.}

\subsection{Interesting information}
\label{sec:interesting_info}

A node considers a received CTP interesting when its body contains at least a
new route, i.e. a route that the node didn't previously know.
In other words, if a CTP contains routes already known by the node, it is
considered uninteresting.

When a node receives an interesting CTP, it forwards the packet to all its
neighbours, excepting the one from which it has received the CTP.
If, instead, the CTP is uninteresting, it will drop the packet.

Note that if a CTP is uninteresting for the node $N$, then it is also
uninteresting for all the other nodes. This is because an uninteresting CTP
contains routes which has been previously received, memorised and forwarded
by the node $N$. Therefore all the other nodes already know the same
routes too.

\subsubsection*{Example}
\label{sec:cycle_qv2_example}

Consider this graph.
\begin{figure}[h]
	\begin{center}
		\includegraphics[scale=0.4]{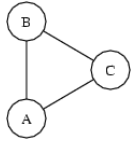}
	\end{center}
	\caption{The A-B-C-A cycle}
	\label{fig:A-B-C-A}
\end{figure}
Suppose that $A$ sends a CTP. 
The two CTPs, after having covered the following two paths will stop:
\[
A \rightarrow B \rightarrow C \rightarrow A \rightarrow B \rightarrow C
\]
\[
A \rightarrow C \rightarrow B\rightarrow A \rightarrow C \rightarrow B
\]
Let's analyze the first CTP step by step, considering that before $A$ sent the
CTP, none knew any route.
\begin{description}
	\item[A $\rightarrow$ B] At this point $B$ doesn't know any route to
		reach $A$, therefore it considers this CTP as interesting and
		forwards it to $C$.
	\item[A $\rightarrow$ B $\rightarrow$ C] By looking at this packet $C$
		learns a route to reach $B$ and $A$.
	\item[A $\rightarrow$ B $\rightarrow$ C $\rightarrow$ A] The node $A$
		learns a route to reach $C$.
	\item[A $\rightarrow$ B $\rightarrow$ C $\rightarrow$ A $\rightarrow$
		B] The node $B$ learns a route to reach $C$.
	\item[A $\rightarrow$ B $\rightarrow$ C $\rightarrow$ A $\rightarrow$
		B $\rightarrow$ C] Finally, $C$ drops the packet, because it
		already knows all the routes contained in it.
\end{description}

From this example we can derive a general result: a CTP will always terminate
in a cycle.

\subsection{Live routes simplification}
The QSPN v2 is the ``live'' version of the Simplify Route algorithm (section
\ref{sec:simplify_routes}).

The CTP flood of the QSPN v2 explores the entire graph, but unlike the ATP
(section \ref{sec:acyclic_TP_flood}), it drops the TPs which contains
redundant routes, thus only the simplified, non redundant routes survives and
continue to explore the graph.

\newpage
\subsection{Cyclicity}
\label{sec:CTP_ciclicity}
When a CTP reaches the extremity of a segment, it is back forwarded, thus it's
as if the extreme nodes had a link with themselves.

\begin{figure}[h]
	\begin{center}
		\includegraphics[scale=0.4]{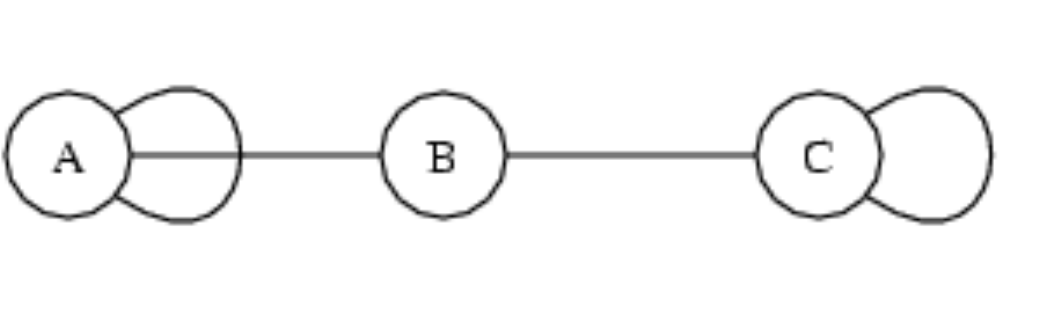}
	\end{center}
	\caption{A segment as viewed from a CTP}
	\label{fig:CTP_segment}
\end{figure}

From the point of view of a CTP, even a segment is a cycle, therefore, for a
CTP, any connected graph is formed just by cycles.

For this reason, a CTP will explore any combination of cycles of the graph.

\subsubsection{Subcycles examples}
These examples highlights some subcycles of a simple graph.

\begin{figure}[h]
	\begin{center}
		\includegraphics[scale=0.4]{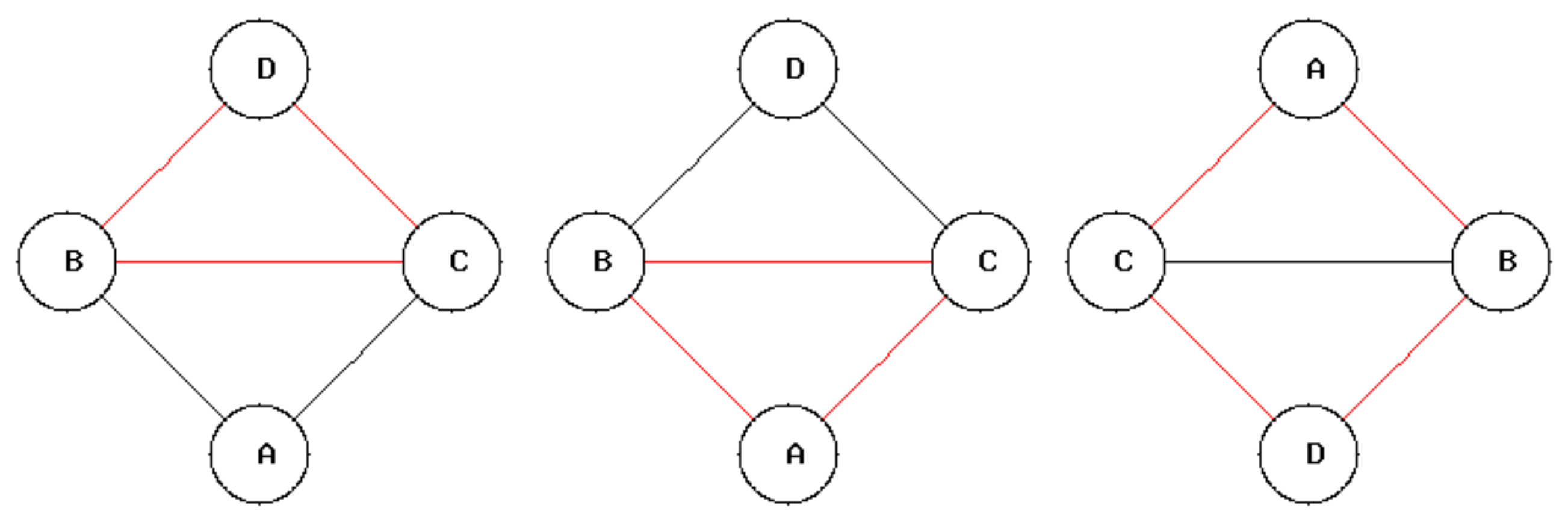}
	\end{center}
\end{figure}
A CTP would explore all these cycles.

\subsection{Finiteness}
$Q^2$ will finish the exploration of the graph in a finite amount of time,
i.e. the flood will terminate.

As we've seen in the example \ref{sec:cycle_qv2_example}, a CTP flood of a
cycle will always terminate. Moreover in section \ref{sec:CTP_ciclicity} we've
noticed that, from the point of view of a CTP, any connected graph is formed
by a combination of cycles. Therefore, a CTP flood of a graph will always
terminate in a finite amount of time.

\subsection{Routes limit}
\label{sec:routes_limit}

Even if $Q^2$ is finite, it still generates too many routes and packets.
Therefore we need to limit the exploration of the graph.\\
An efficient and elegant solution is to further define what the ``interesting
information'' is:

Let each node of the network keep a maximum of \emph{MaxRoutes} routes in
its memory. A node considers a received CTP interesting when its body contains
at least a route which is more efficient than the previously memorised routes.
The efficiency of a route can be quantified with a convenient parameter, f.e.
the rtt or the bandwidth capacity.
If the node has reached the \emph{MaxRoutes} limit, it will substitute the old
route with the more efficient one.

Note that this definition is more general than the previous. Indeed, if the
node $S$ doesn't know the route to reach $D$, the efficiency of the route $S
\rightarrow D$ is equal to $0$.

A node can also keep in memory more than \emph{MaxRoutes} routes, because this
limit applies only to the number of routes which will be used to evaluate the
received CTP.

\subsection{Scalability}
\label{sec:Scalability}
We will now exploit the bouquet property of the Tracer Packets, which has
been described in section \ref{sec:proprieties_TP}.

Suppose that in our network every link has the same bandwidth capacity and
that the generic node $D$ doesn't know any route to reach the
node $T$.
If a TP, received by the node $D$, contains a new route $t$ that
connects $D \rightarrow T$, then we can deduce, by the bouquet property, that
$t$ is the fastest route between $D$ and $T$.

The immediate consequence is that $D$ will receive all the other $D
\rightarrow T$ routes in order of efficiency: the first is, as we've seen,
the best route, the second one will be slower than the first but surely
better than any other, and so on.

\subsubsection{TP Classes}
\label{sec:tp_classes}
Two different routes can be very similar, because they can differ only in a
small part. Two routes which differs of just one hop, are almost identical.
For this reason, other than the best $D \rightarrow  T$ route, the CTP will
also explore all the other routes which are almost identical to it.

We can thus order all the TPs which $D$ will receive into classes. The first
class, denoted with $[1]$, contains the TP which have covered the best
route and all the others similar to it, i.e. all the other routes with the same
number of hop and a similar trtt (total round trip time).
The second class $[2]$, contains the TP which have covered routes which are
less efficient than those contained in $[1]$ but are more efficient than those
of class $[3]$.
More generally we can say that the $[n\textrm{-th}]$ class contains the routes that,
if included in a list of all the routes of the graph, ordered in decrescent
order of efficiency, will be listed starting from the position $(n-1)c+1$. Where
$c$ is the numbers of routes contained in each class.

In the classes we are including \emph{routes} and not \emph{tracer packets},
because a TP may contain more than one route.

\subsubsection{Subcycle filter}
Each node of the graph acts as a filter for all the subcycles containing it.

Suppose the node $D$ is contained in the subcycle $\sigma$ and that a CTP $t$
enters in it (through another node).
If the node $D$ has the \emph{MaxRoutes} limit set, it will memorise, for each
node of the network, only the first \emph{MaxRoutes} received routes, while the
rest will be disregarded, and \textbf{not} forwarded. \\
The CTP $t$ won't be forwarded by the node $D$, if it contains routes which
exceed the \emph{MaxRoutes} limit, but this is true for all the nodes of
$\sigma$, therefore $t$ won't even be able to escape from the subcycle
$\sigma$. However, this also means that all the CTPs, which are in a higher
class than that of $t$, won't be allowed to pass from $\sigma$.\\

Since this happens for all the subcycles of the graph, we can conclude that at
worst, the number of CTPs increases in polynomial time with the increase of
subcycles.

\subsubsection*{Example}
Consider this graph.
\begin{figure}[h]
	\begin{center}
		\includegraphics[scale=0.5]{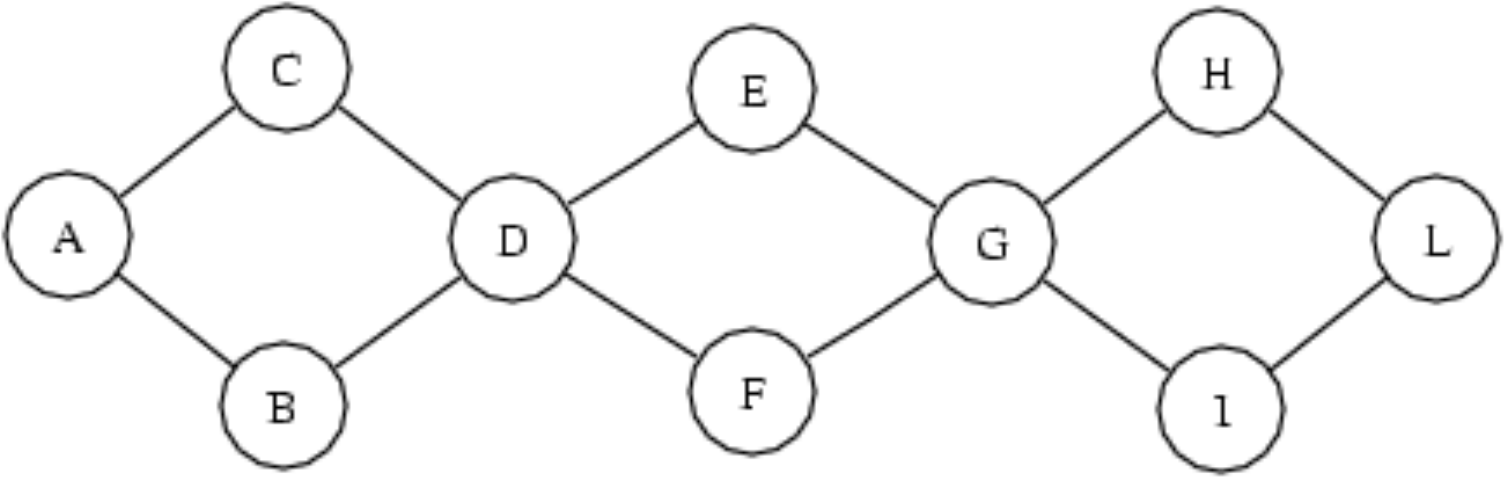}
	\end{center}
	\caption{Sequential composition of cycles}
	\label{fig:q2_scalability_example}
\end{figure}
Let's analyse the difference between a CTP without limits and the same CTP
with \emph{MaxRoutes} set.
\begin{description}
	\item[Unlimited CTP] $A$ sends a CTP. Suppose for simplicity that
		this CTP won't loop inside a cycle.\\
		$D$ receives $2$ packets from $A$, $G$ $2*2$ from $D$ and
		finally $L$ $2*2*2$ from $D$.\\
		$L$ sends back all the received CTPs, thus $G$ gets $2^4$ CTPs
		from $L$, $D$ gets $2^5$ and finally $A$ receives $2^6=64$
		packets.
		Obviously this is too much for this simple graph.
	\item[Limited CTP] Let's use $\textrm{\emph{MaxRoutes}}=1$ and suppose
		that the CTP won't loop inside a cycle. We'll write
		$D_A(n,z)$ to indicate that $D$ received $n$ CTPs from the node
		$A$ but has kept and forwarded only $z$ packets.\\
		$A$ sends a CTP.\\
		\begin{align*}
			&A_{\emptyset}(0,2) \rightarrow	D_A(2,2)\rightarrow
			G_D(4,3) \rightarrow L_G(6,4) \rightarrow \\
			& \qquad G_L(8,5) \rightarrow D_G(10, 6) \rightarrow
			A_D(12, 0)
		\end{align*}
		At the end, A gets $12$ packets.\\
		Each node forwards $p/2+1$ packets, where $p$ is the number of
		received CTPs. This is because the first two packets give to
		the node a new route, while the other two, and the successive
		ones, cover a superflous route. For example, consider
		\begin{align}	
			&\underline{A \rightarrow  C \rightarrow D \rightarrow
			E \rightarrow  G} \label{eq:d1}\\
			&A \rightarrow  C \rightarrow D \rightarrow
			\underline{F \rightarrow  G} \label{eq:d2}\\
			&A \rightarrow  \underline{B \rightarrow D} \rightarrow
			E \rightarrow  G \label{eq:d3}\\
			&A \rightarrow B \rightarrow D \rightarrow F
			\rightarrow  G \mathbb{!}\label{eq:d4}
		\end{align}
		The underlined routes are the new route for $G$. As you can
		see, in the CTP (10) G doesn't find any new route, so
		it drops the packet and doesn't forward it.
	\end{description}

\subsubsection{Efficiency order}
\label{sec:eff_order}

We've noted in section \ref{sec:Scalability} that any node will receive the
CTPs in order of efficiency, thus we are sure that only the first
\emph{MaxRoutes} received routes (which point to a specific node) are
meaningfull, while all the successive ones are uninteresting and should be
dropped. This is important, because each
node will first receive \emph{MaxRoutes} interesting routes and
\textbf{then} the uninteresting ones, that will be dropped. 
Doing so, an uninteresting route won't be forwarded before its interesting
correspondent, and as soon it is recognised it will be dropped.

To clarify this concept suppose that the routes aren't received in order of
efficiency and that $\textrm{\emph{MaxRoutes}}=2$. Then suppose that node $D$ doesn't
have any route yet and receives a CTP of class $[7]$. $D$ will consider this
CTP as interesting and forward it, because it's the first one it receives.\\
In conclusion, the CTP of class $[7]$ would be allowed to be propagated among
the network even if the \emph{MaxRoutes} has been set to $2$ routes.
Instead, if $D$ receives the packet in an ordered
manner, it will first get the CTPs of class $[1]$, then those of class $[2]$
and so on. For this reason the CTP of class $[7]$. and all the other which
exceed the \emph{MaxRoutes} limit, won't be allowed to pass from $D$.

\subsection{Bandwidth issues}
Until now we've supposed that every link of the network has the same bandwidth
capacity. However, we'll see in section \ref{sec:bandwidth_q1q2} that the QSPN
can be used in the general case.

\subsection{Worst case}
\label{worstcase}
The graph, formed by $n$ nodes, which has the maximum number of cycles is the
worst case for $Q^2$, because, if not limited, it will have to explore any
combination of cycles.

Such graph is the \emph{complete graph} \cite{completegraph}, and the total
number of its subcycles is:
\[
\sum_{k=3}^n\,\frac{1}{2}\binom{n}{k}(k-1)!
\]

\section{QSPN v1}
\label{sec:QSPNv1}
\small{It isn't necessary to read this paragraph in order to understand the
rest of the paper. If you aren't interesent in the QSPN v1, just skip over.}

The QSPN v1 is a restricted case of $Q^2$. 
It is divided in two phases.
The first one is called qspn\_close:
a node sends a QTP (QSPN Tracer Packet) called qclose, this node becomes a qspn starter.
A qclose is a modified form of tracer packet.
  A node $N$, which receives a qclose from the link $l$, marks as ``closed'' the
  same link $l$ and forwards the packet to all its other neighbours.
  All the following qclose packets received by the same node $N$, will
  be forwarded only to the links which have not been already closed.\\
  During the graph exploration, some nodes will close all their links.
  These nodes are called \emph{extreme nodes}. When a node becomes an extreme node, 
  it will send another type of tracer packet, called qspn\_open (which is also
  the name of the second phase) to all its neighbours, except the one from
  which it received the last qclose packet (let's call this neighbour $L$).
  The qopen packet sent to $L$ is empty, while those sent to the other
  neighbours contains the body of the last received qclose packet.\\
  The qopen behaves as the qclose: it "opens" the links, however the nodes
  which have all their links opened won't forward any other packets.\\
\textbf{Example: }
  \begin{figure}[h]
  \begin{center}
	  \includegraphics[scale=0.4]{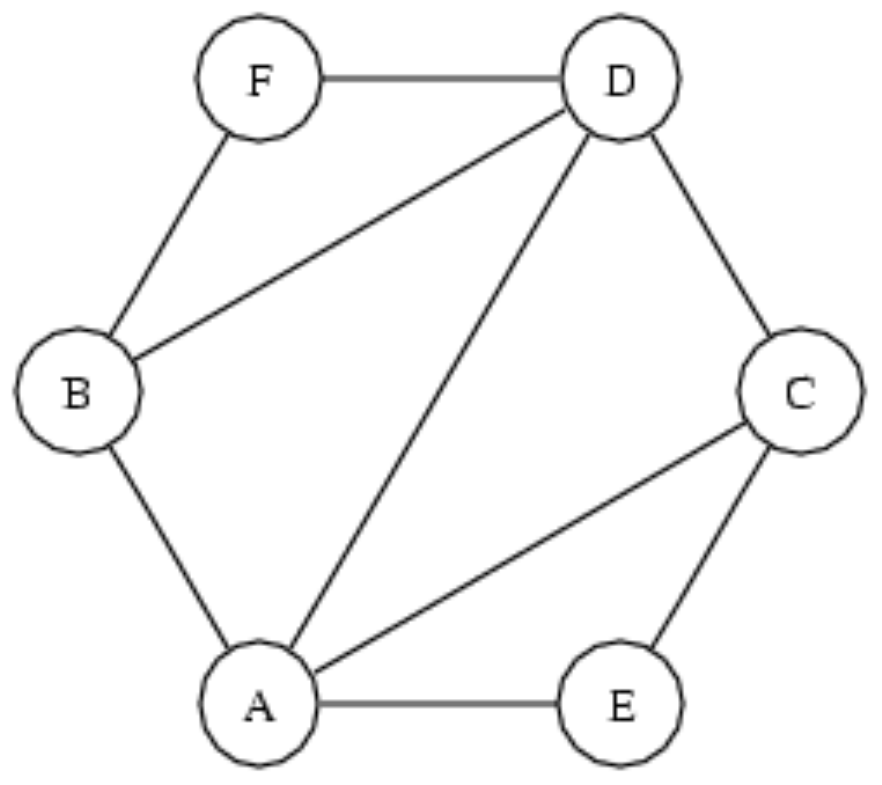}
	  \caption{Example of a QSPN v1 graph exploration}
	  \label{fig:q1ex}
  \end{center}
  \end{figure}
  Consider figure \ref{fig:q1ex}.
  \begin{itemize}
	  \item The node $E$ sends a qclose. It is now a qspn starter.
	  \item Suppose that the node A receives the qclose before $C$. $A$
		  closes the link $E\rightarrow A$ and forwards the qclose to
		  $B$, $C$ and $D$.
	  \item $C$ receives the qclose from $E$, closes the link $E\rightarrow C$ and forwards it to A and $D$.
	  \item $C$ receives the qclose from $A$ and closes the link.
	  \item $B$ and $D$ have received from $A$ the qclose and close the respective links.
	  \item Suppose that $B$ is the first to forwards the qclose to $F$.
	  \item $D$ forwards the qclose to $F$, but at the same time $F$ forwards it to $D$.
	  \item $D$ receives the qclose from $B$, too.
	  \item $D$ and $F$ have all the links closed. They send a qopen.
	  \item The qopen propagates itself in the opposite sense.
	  \item The qopen ends.  Each node has the routes to reach all the other nodes.
\end{itemize}

The qspn\_close phase can be seen
as a CTP with the added rule that when two CTPs collide, they will be
converted to two normal TPs (the qopen phase).

\subsection{$\mathbf{Q}$ vs $\mathbf{Q^2}$}
These are the substantial differences between Q (QSPN v1) and $Q^2$ (QSPN v2):
\begin{enumerate}
	\item Q generates less packets than $Q^2$, because in the qopen
		phase it uses normal TPs which expires quickly. The side
		effect of this behaviour is that Q may not discover all
		the best routes. However Q gives at least one route to
		reach each node of the graph.

	\item Q uses less memory than $Q^2$, because it just keeps a
		forwarding table, instead, $Q^2$ needs to memorize 
		$\emph{MaxRoutes}$ complete routes to evaluate the successive
		CTPs. By the way, this difference is minimal.

	\item $Q^2$ doesn't need synchronization. the CTPs doesn't need to
		have an ID, thus many nodes can send simultaneously or
		asynchronously a CTP without creating any problem.\\
		This isn't the same in Q, which requires a strict
		synchronization between the nodes: two nodes can send 
		a qclose only at the same time.
	
	\item This is a consequence of the propriety described above: every
		time a node joins the net or dies or its rtt/bw capacity
		changes, it is possible to immediately send a CTP. Indeed, if
		the changes in the local gnode regard that node only, the CTP
		will be like a normal Tracer Packet (see \ref{sec:TP_flood}).

	\item $Q^2$ is easier and simpler than Q to be implemented. In
		general this means that the code of $Q^2$ will have less
		bugs.
\end{enumerate}

From this comparison we can conclude that $Q^2$ is preferable over Q.

\section{Network dynamics}
\label{sec:netdyn}
The QSPN v2 defined until now is not suitable for dynamic networks. As
example, consider this problem:

suppose that the whole graph has been already explored, and thus every node
has at least one route to reach all the other nodes. Consider the case when
the efficiency of a link, f.e. $N \leftrightarrow P$, worsens.
$N$, in order to update the maps of the other nodes, sends a CTP to its
neighbour $P$, and $P$ forwards it to its neighbours. However this CTP will be
immediately dropped! Indeed, the nodes will consider this CTP not interesting,
because the contained $\dots\rightarrow P \rightarrow N$ route is less
efficient than the old one, which has been saved during the last graph
exploration.

\subsection{Extended Tracer Packet}
\label{sec:etp}
The ETP solves the problem of how the graph should be re-explored to
update the maps of the nodes interested to a network change.
Its way of working is based on a simple observation:

The first QSPN exploration distributes, among the nodes, information describing
the network topology. 
When a change in the network occurs, only
the information stored in the nodes affected by the change must be updated.
The unaffected nodes will still have up to date information that they can simply
redistribute with the use of the Extended Tracer Packets.

An ETP is an Acyclic Tracer Packet,\footnote{
An ATP (see paragraph \ref{sec:acyclic_TP_flood}) is a normal TP with the following
rule: a node drops the received ATP if  its  node ID is already present in the route
contained in the body of the packet. Note also, that since it is a normal TP,
it is not reflected back, when it reaches the end of a segment.}
which contains a portion of a map. Since a map is a set of routes and a
route can be described by a TP, the ETP can be considered as different TPs packed togheter.
Each TP of the ETP is then subjected to the rules of the QSPN v2.

In order to give an exact definition of an ETP, we must examine each case of network change.
\begin{description}
	\item[Worsened link] 
		\label{wlink}
		Suppose that the link $A
		\stackrel{l}{\leftrightarrow} B$ worsens.\\
		Let's analyse what $B$ will do (the situation is symmetric for
		$A$).

		$B$, if interested in the change, will create an ETP
		containing all its old routes, affected by the change, and all
		the backup routes used as substitute for the old ones. The ETP
		will be sent to all its neighbours, except $A$. In detail,
		$B$ will use the following algorithm:
		\begin{enumerate}
		\item If at least one of the primary
		       routes\footnote{
				a route src$\rightarrow dst$ is called primary
				if it is among the first MaxRoutes routes of
				type src$\rightarrow dst$}
			saved in the map
			of $B$ and different from the route $B\rightarrow A$, uses the link $l$, then $B$ creates an
			empty ETP, otherwise the algorithm halts, i.e. $B$
			won't do anything.
		\item \label{upmap}
			If the empty ETP has been created, $B$ updates its maps:
			suppose that the route $r$, passing through the link
			$l$, had a total rtt $t_0(r)$. If the rtt of the link
			$l$ before the change was $t_0(l)$ and now is
			$t_1(l)$, where $t_0(l) < t_1(l)$, then
			$t_1(r):=t_0(r)-t_0(l)+t_1(l)$.\\
			For the bandwidth we have:
			\begin{align*}
				&b_0(r)\;\;\T{ the total bandwidth of the
				route $r$, before the change}\\
				&b_0(l) > b_1(l)\;\;\T{the bw of $l$ has
				worsend during the change}\\
				&b_1(r):=\min \{b_0(r), b_1(l)\}
			\end{align*}
			The routes are then sorted.
		\item \label{stepR} 
			$B$ creates the temporary set $Q$, 
			containing all the primary
			routes passing through the link $l$. From $Q$
			it creates the set $R$, where
			\[
			R=\{r\in M\;|\;\exists q\in Q:\;\T{dst}(r)=\T{dst}(q)\}
			\]
			where $M$ is the set of all the primary routes of the map, and 
			$\T{dst}(r)$ is the destination of the route $r$. In
			other words, $R$ is the set of primary routes having
			the same destination of at least one route of $Q$.
			Note that $Q\subseteq R$.
			Each
			route $r\in R$ is saved as $(\T{dst}(r), \T{rem}(r), \T{tpmask}(r))$, where $\T{rem}(r)$ is the Route
			Efficiency Measure, and $\T{tpmask}(r)$ is a bitmask of 256
			bits, where the bit at the i-th position indicates if
			the node i is an hop of the route $r$.
		\item $B$ fills the ETP: 
			\begin{enumerate}
				\item it adds in it the set $R$
				\item it appends the ID of $A$, along with the efficiency
					value of the link $l$, and, as usual, its ID.
				\item it sets to 1 the \emph{flag of interest}.
			\end{enumerate}
		\item Finally, $B$ sends the ETP to all its neighbours, except $A$.
		\end{enumerate}
		\label{ETPrule1}
		Suppose that the neighbor $C$ of $B$ has received the ETP. $C$
		will examine the ETP and, if considered interesting, it 
	will update its map and forward the ETP to the other neighbours, as
		follow:
		\begin{enumerate}
		\item If the ID of the node $C$ is already present in the
			received ETP, then $C$ immediately drops the ETP, and
			skips all the following steps\footnote{This is the acyclic
			rule}.
		\item Let $R$ be the set of routes contained in the
			ETP received by $C$.\\
			Let $M$ be the set of all primary routes contained in the map of $C$.\\
		For each route $r\in R$, the node $C$ looks for 
		a route $m\in M$ such that 
		\[\T{dst}(m)=\T{dst}(r),\;\;\T{tpmask}(m)=\T{tpmask}(r)\]
		If $m$ exists, then $C$ sets $\T{rem}(m):=\T{rem}(r)$
		\footnote{with this operation we are actually replacing $m$
		with $r$, in the map $M$}.
		Otherwise, $r$ is copied in the temporary set $R'$.\\
		$M$ is sorted, i.e. the routes of the map of $C$ are sorted in
		order of efficiency.
	\item For all $r' \in R'$,
		\begin{enumerate}
			\item if $r'$ is a better alternative to at least one primary route
				$m'\in M$ such that $\T{dst}(r')=\T{dst}(m')$,
				then $r'$  is saved in the map of $C$
				(note \footnote{When saving a route $r$ of the ETP
				in the map, we must consider the hops covered by the ETP,
				so the real saved route is
				$r\leftarrow \T{hop}_1\leftarrow \T{hop}_2\leftarrow
				\dots\leftarrow \T{hop}_n$. In this
				case we'll have $r\leftarrow B\leftarrow A$.}), 
			\item otherwise, $r'$ is removed from $R$.
		\end{enumerate}
		Note \footnote{this step implements the QSPN v2 rules: only
		good routes are kept, the other are discarded. Notice
		the extension: if the ETP had only one route, it would be
		almost equal to a CTP (the CTP doesn't have the acyclic rule)}
	\item If $R$ is now empty, i.e. all its routes have been removed, then
		$C$ considers the ETP as uninteresting. Let's suppose for now
		that it is interesting: the \emph{flag of interest} remains
		set to	1.
	\item $C$ packs the ETP with the previously modified
		set $R$, and adds its ID. The ETP is sent to all its neighbours,
		except $B$
	\end{enumerate}
	The rnodes of $C$ will use this same procedure. In this way, the ETP
	will continue to be propagated until it is considered interesting.

	Let's suppose now that a node $N$ receives the ETP and considers it
	uninteresting. $N$ won't just drop the ETP, but will also send back
	another ETP containing its own routes. The reason is simple: $N$
	considers the received ETP uninteresting, this means that $N$ isn't
	affected by the change of the link $l$, i.e. all its primary routes don't
	pass through $l$ and thus are still optimal. Therefore, $N$ will
	send back its routes, hoping that they will be useful to the nodes
	affected by the change. In detail, this is what will happen when $N$
	receives the ETP:
	\begin{enumerate}
		\item $N$ receives the ETP from the node $L$, and considers it
			uninteresting.
		\item Let $R$ be the set of routes contained in the ETP.\\
		Let $M$ be the set of all primary routes contained in the map
		of $N$.\\
		$N$ creates the following set:
		\[
		S=\{m\in M\;|\; \exists r\in R:\;\T{dst}(m)=\T{dst}(r)\}
		\]
		\item $N$ creates the new ETP, appending in it the set $S$ and
			its ID. The \emph{flag of interest} of this ETP is set
			to 0.
		\item The ETP is sent to $L$.
	\end{enumerate}
	At this point, the new ETP created by $N$, will propagate back in the
	same fashion of the previous ETP (see page \pageref{ETPrule1}), i.e.
	until considered interesting. The only difference is that when a node
	considers it uninteresting, it is just dropped\footnote{the node will
	drop the ETP if the \emph{flag of interest} is set to 0 and if it is
	uninteresting}.
	\item[A node dies]
		Suppose the node $A$ dies. Each neighbour $B$ of $A$ will send
		an ETP. The ETPs are generated and propagated with the
		algorithms described in the \textbf{worsened link} case (page
		\pageref{wlink}), the differences are:
		\begin{enumerate}
			\item Instead of considering the routes passing
				through the worsened link we consider the
				routes passing from the dead node.
			\item Suppose that the node $N$ receives the ETP from
				its neighbour $L$ and considers it uninteresting.
				$N$ sends back to $L$ the new ETP to share its
				routes
				among the interested node. However, unlike the
				case for the worsened link, $N$ creates also a
				new simple TP, where it writes only the
				information of the death of $A$.
				$N$ sends this TP to all its neighbours, except $L$.

				This simple TP will be propagated with the
				rules of the QSPN v2. For this reason and since it carries
				only one useful information (the death of
				$A$), each node will receive it just once (the
				second time it will be dropped). This simple
				TP serves to inform the nodes, unaffected
				by the network change, of the death of $A$.
		\end{enumerate}
	\item[Improved link]
	\label{ilink}
		Suppose that the link $A
		\stackrel{l}{\leftrightarrow} B$ improves.\\
		Let's examine the events, starting from $A$, keeping in mind
		that the situation is symmetric for $B$.

		Since the link $l$ improved, it may be possible for $B$ to use
		it to improve some of its routes. For this reason, $A$ will
		send to it an ETP with all the routes of its map, except those of the form
		$A\rightarrow B\rightarrow \rightarrow \dots$. If $B$ finds
		something of interest, it will forwards the ETP. In detail:
		\begin{enumerate}
			\item Let $M$ be the set of all primary routes
				contained in the map of $A$.\\
			      $A$ creates the following set:
			      \[
			      R=\{m\in M\;|\; \T{gw}(m)\neq B\}
			      \]
			      where $\T{gw}(m)$ is the first hop of the route
			      $m$. 
			      Each route $r\in R$ is saved as $(\T{dst}(r),
			      \T{rem}(r), \T{tpmask}(r))$.
		      \item $A$ creates the ETP:
			\begin{enumerate}
				\item it writes in it the set $R$
				\item it appends the its node ID, along with the efficiency
					value of the link $l$.
				\item it sets to 1 the \emph{flag of interest}.
			\end{enumerate}
		      \item It sends the ETP to $B$
		\end{enumerate}
		At this point, the ETP is propagated exactly in the same way
		of the \textbf{worsened link} case (see page \pageref{wlink}).
	\item[A new node joins]
		Suppose the node $A$ is joining the network. Its neighbours
		are $B_1, B_2,\dots, B_n$, which are all already hooked, i.e.
		they aren't joining. Then,
		\begin{enumerate}
			\item Each neighbour $B_i$ sends its whole map to
				$A$
			\item $A$ waits until the maps of all its neighbours
				are received.
			\item The maps are ``merged'' into a single map, which
				becomes the map of $A$. In simple words, the
				merge of two maps result in a map having only the
				best routes of the two.
			\item If the neighbours of $A$ are more than one, i.e.
				$n>1$, then $A$ sends, to each of them, an ETP
				containing all the primary routes of its map.
			\item The ETPs are propagated exactly in the same way
				of the \textbf{worsened link} case (see page
				\pageref{wlink}).
		\end{enumerate}
		Note that this case extends the hooking procedure.
	\item[Broken link] 
		The case where the link $A
		\stackrel{l}{\leftrightarrow} B$ becomes invalid, is handled
		in the same way of the \textbf{worsened link} case (see
		page \pageref{wlink}), because we can consider $l$ as
		infinitely worsened.
	\item[New link]
		The case where a new link $A \stackrel{l}{\leftrightarrow} B$
		is established between, is handled  in the same way of the
		\textbf{improved link} case (see page \pageref{ilink}),
		because we can consider $l$ as infinitely improved.
\end{description}
%

\section{QSPN optimisations}
\subsection{Rtt and bandwidth}
\label{sec:bandwidth_q1q2}

The bandwith capacity of a route can be used as a parameter of its
efficiency. In this section we'll analyse the implications for the QSPN.
For more information about the bandwidth management in Netsukuku you can read
the NTK\_RFC 002 \cite{ntkrfc0002}.

\subsubsection{Rtt delay}
\label{sec:rtt_delay}

Each node of the network will delay the forwarding of a received CTP by a time
inversely proportional to its upload bandwidth. In this way the CTPs will
continue to be received in order of efficiency (see section
\ref{sec:eff_order}).
The side effect of this rule is that the extreme cases will be ignored, i.e. a
route with a very low rtt but with a very poor bw, or a route with an optimal
bw but with a very high rtt. However, in the ``real world'' these extreme
cases are rare, because the rtt and the bw are often related.

\subsubsection{Asymmetry in $Q^2$}
The QSPN v2 is a very flexible algorithm that can be adapted to a large range
of cases. Indeed, with a minimal added overhead, it is possible to achieve
asymmetric routing discovery, i.e. a discovery that discerns the upload
bandwidth of a route from its download one.\\
We call this extension the \emph{asymmetric QSPN v2}, while we refer to the
old one as \emph{symmetric $Q^2$}.

\begin{enumerate}
	\item	First of all, it is necessary to define further the ``interesting
		information''. A CTP will be considered interesting, not only when it contains
		interesting (see \ref{sec:interesting_info}) download routes, but also upload
		ones. In other words, we consider the upload sense of a route too. 

		For example, suppose that the node $A$ received the CTP $ABCDAERTA$. In this
		case $A$ will know two dinstinct upload routes: $A\rightarrow B\rightarrow
		C\rightarrow D$ and $A\rightarrow E\rightarrow R\rightarrow T$. Instead, in
		the classic CTP, $A$ would have known only $A\rightarrow T\rightarrow
		R\rightarrow E$ and $A\rightarrow D\rightarrow C\rightarrow B$.
	\item Secondly, since we are considering the reverse (upload) routes
		too, we have to remove the restriction imposed on the CTP,
		which has been described in section \ref{sec:reflected_CTP}.
                The body of the CTP reflected from the extreme of a segment
                won't be erased, thus it will contain the old routes too.
		This is because the old routes can contain interesting
		information about upload routes.
		For example, consider this segment:
		\[ \cdots \leftrightarrow A \leftrightarrow B \leftrightarrow C \leftrightarrow N \]
		If $N$ doesn't erase the route received in the CTP, $A$ will
		receive the following CTP:
		\[
		\cdots \rightarrow A \rightarrow B \rightarrow C \rightarrow N \rightarrow C \rightarrow B \rightarrow A
		\]
		In this case $A$ will know the following upload route:
		\[
			A \rightarrow B \rightarrow C \rightarrow N
		\]
		When parsing a CTP, a node will recognize the part of the
		routes which are in the form of $XacaY$, where $a$, $c$ are
		two nodes and $X$ and $Y$ are two generic routes. The packet
		will then be split in $Xac$ and $caY$.
\end{enumerate}
At this point we've finished. In fact, we are sure to receive at least one
upload route per node because a CTP traverses each path first in one sense
and then in the opposite. The CTP information filter, will allow us
to receive only the best routes. However, since the Rtt Delay
(\ref{sec:rtt_delay}) is tuned for download routes only, it is possible that
some upload paths will be ignored.\\
It is interesting to note that in the majority of cases, the number of
CTPs will remain equal to that of the symmetric $Q^2$.

\subsection{Disjoint routes}
The routing table of each node should be differentiated, i.e. it should not
contain redundant routes.

For example, consider these $S \rightarrow D$ routes:
\begin{align}
	& SBCFG_1G_2G_3G_4G_5G_6G_7 \dots G_{19} D	\\
	& SRTEG_1G_2G_3G_4G_5G_6G_7 \dots G_{19} D	\\
	& SZXMNO_1O_2O_3O_4O_5D				\\
	& SQPVY_1Y_2Y_3Y_4D
\end{align}
The first two are almost identical, indeed they differ only in the first three
hops. The last two are, instead, totally different from all the others.\\
Since the first two routes are redundant, the node $S$ should keep in memory only
one of them, saving up space for the others non-redundant routes.
\newline

Keeping redundant routes in the routing table isn't optimal, because if one of the
routes fails, then there's a high probability that all the other redundant
routes will fail too. Moreover when implementing the multipath routing to load
balance the traffic there won't be any significative improvements.
\newline

$Q^2$ itself should avoid to spread redundant routes. In order to achieve
this result, we refine the efficiency value associated to a route. Suppose we
want to affect the efficiency value $R_e$ assigned to the route $R$:
\begin{enumerate}
	\item let $0\le s(R,S)\le 1$ be the similarity level of the route $R$
		with $S$.
	\item for each memorised route $S$ we compute $s(R,S)$ and if we find
		a route $S$ such that $s(R,S) > 0.5$ we go to step 3.
	\item we set \[R_e = R_e\frac{1-s(R,S)}{k}\] where $k$ is an appropriate
		coefficient.
\end{enumerate}
As explained in section \ref{sec:routes_limit} the efficiency of a route is
used as a parameter to evaluate its interest, therefore the more a route is
similar to a memorised route the more its efficiency will decrease. Hence it
will be considered less interesting.\\
Note that this is a generalization of concept of interesting route defined in
section \ref{sec:routes_limit}, in fact, when $R$ and $S$ are equal, we'll
have
\[s(R,S)=1\]
so the $R_e$ value will be equal to $0$.

\subsection{Cryptographic QSPN}
A node could easily forge a TP, injecting in the network false routes and
links information. The attack would just create a temporary local damage,
thanks to the distributed nature of the QSPN. However the optimal solution is
to prevent these attacks.\\

A node, whenever joins Netsukuku, generates a new RSA key pair. The
continuous generation of keys prevents the leakage of the node's anonymity.
The node shares its public key to all the other nodes of its group node.\\
Each entry appended in a TP is then signed with its private key, doing so the
other nodes will be able to prove its authenticity and validate the path
covered by the TP.\\
The size required to store the signatures in the TP can be kept costant using the
aggregate signature system \cite{aggrsign1} \cite{aggrsign2}.

\section{Simulating the QSPN v2}
As a proof of concept, we've a written the q2sim.py\cite{q2sim}, a simulator
implementing the core of the QSPN v2.\\
The q2sim is  an event-oriented Discrete Event Simulator.
Each (event,time) pair is pushed in a priority queue.
The main loop of the program retrieves from the queue the event having the
lowest `time' value. This "popped" event is executed. In this case, the
events are the packets sent on the network.

In this paragraph we'll analyze the results of various simulations.
\begin{description}
	\item[TP flux] The \emph{TP flux} of a node $n$, is the number of
		distinct\footnote{To clarify what we mean with ``distinct
		TPs'': suppose $n$ has $6$ neighbours, it receives only two
		TPs and sends only a new one, then $\Phi(n)\neq 5+5+6$, but
		$\Phi(n)=3$.}
		TP packets which have been forwarded by $n$ during the entire QSPN
		exploration. it is indicated with $\Phi(n)$.
	\item[Mean TP flux] Given $k$ nodes ${n_1,\dots,n_k}$, their mean TP
		flux is:
		\[
		\Phi_m({n_1,\dots,n_k})=\frac{\sum_{i=1}^{k}\Phi(n_i)}{k}
		\]
	\item[Starter node]
		A starter node is a node which sends the first TP in a graph,
		not yet explored by the $Q^2$. There can be more than one
		simultaneous starter node.
\end{description}
In all the following tests the \emph{MaxRoutes} limit has been set to 1.\\

As a general result, the mean TP flux is proportional to the number of
subcycles present in the graph. Thus the maximum mean TP flux is reached in a
complete graph, where each node is connected to all the others. In a complete
graph of $k$ nodes the mean TP flux is approximately:
\[
	\Phi_m \approx k-1,\;\; \Phi_m \le k
\]
The figure below shows this result. In the $x$ axis, an integer point
corresponds to the number of nodes of the complete graph. The adjacent points of the
graph have been connected  with straight line segments.
\begin{center}
\label{fig:k_one_starter}
\includegraphics[scale=0.45]{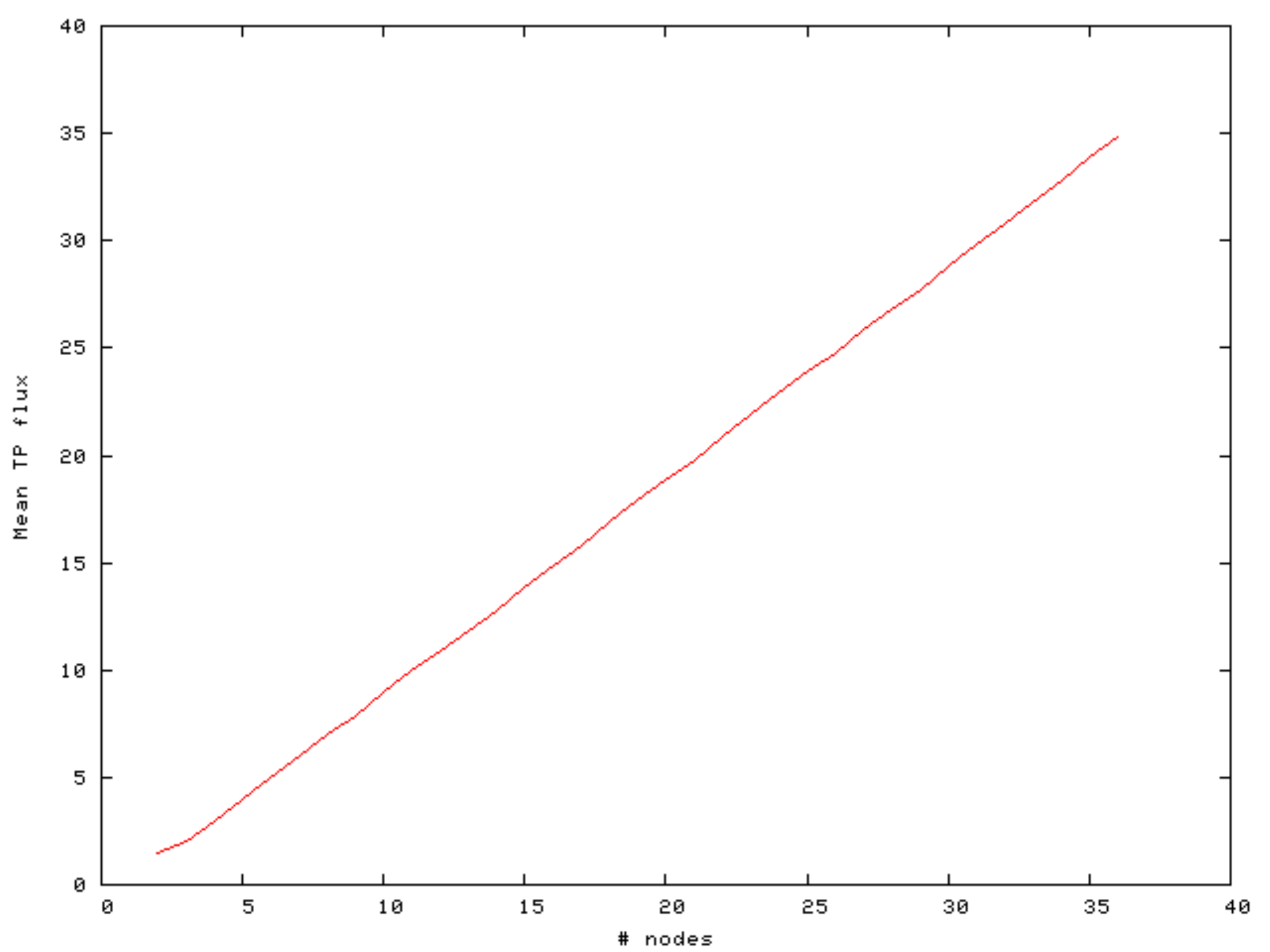}
\end{center}
By increasing the number of starter nodes the
mean TP flux increases slightly until it reaches $\Phi_m=k$. Indeed, if all the
nodes of the graph are starters, then each of them will send a TP to all the
other nodes. The increasing of the number of starter nodes decreases the time required
to complete the exploration of the graph.
The figure below shows this result. In the $x$ axis, an integer point
corresponds to the number of starter nodes. Each plotted line corresponds to a
complete graph with different number of nodes. The first bisector has been
plotted to facilitate the reading of the graph.
\begin{center}
\label{fig:k_starters}
\includegraphics[scale=0.45]{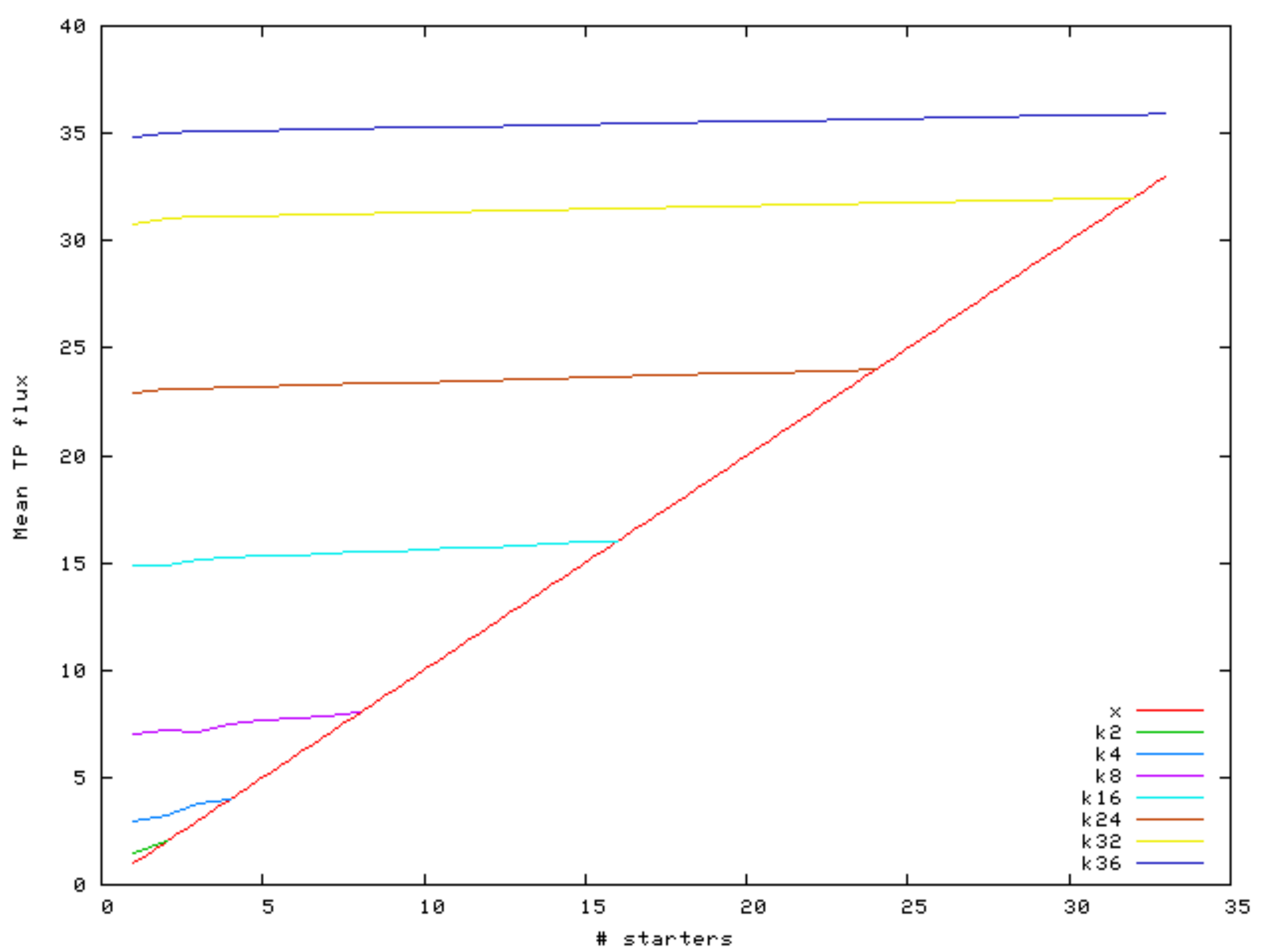}
\end{center}
The complete graph is the worst case for the $Q^2$, therefore in the general case the mean TP flux will be:
\[ \Phi_m \le n \]
where $n$ is the number of nodes of the graph.
For example, in random graphs with increasing number of nodes, the mean TP
flux will assume the distribution shown in the figure below:
\begin{center}
\label{fig:rone_starter}
\includegraphics[scale=0.45]{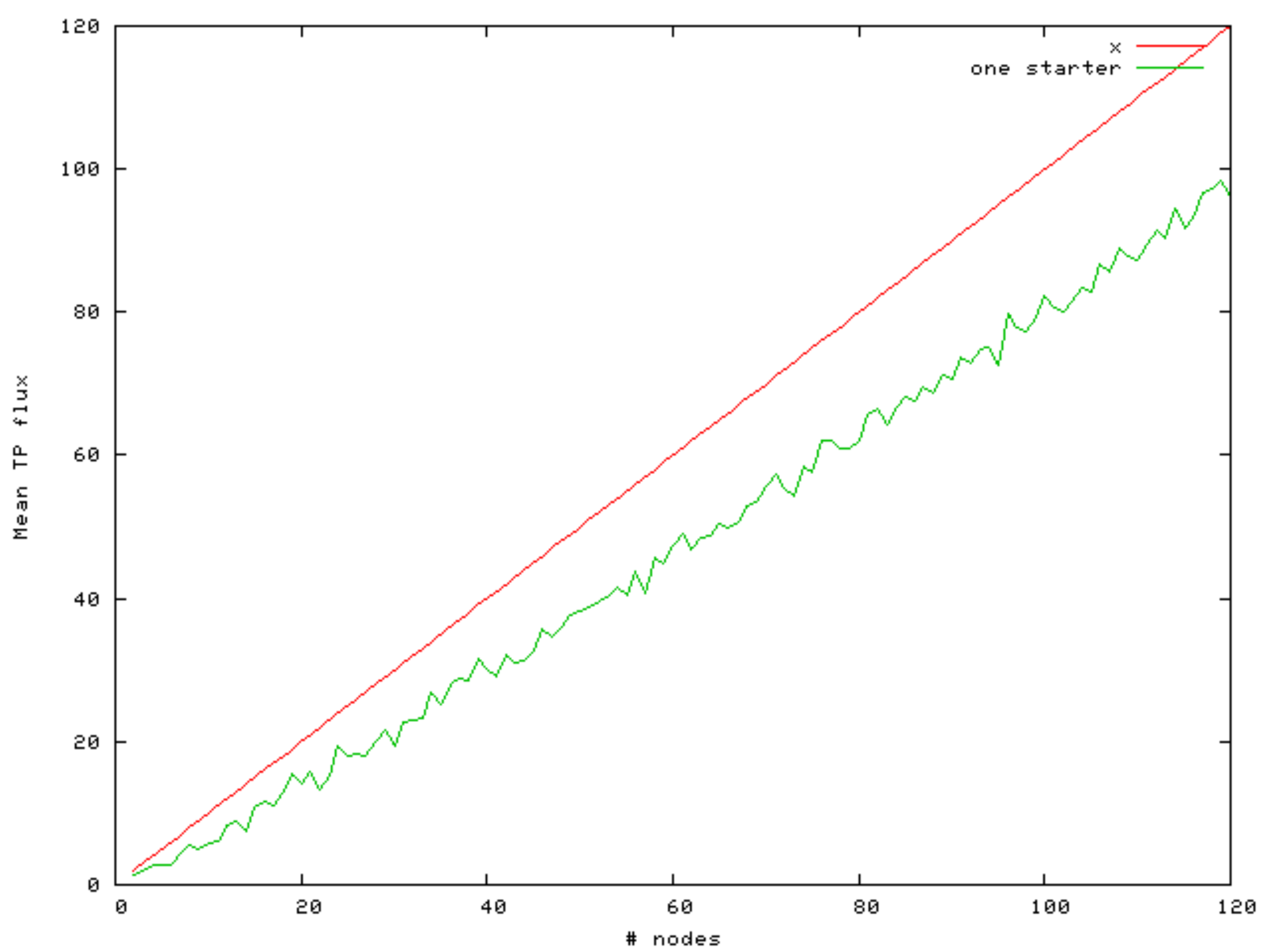}
\end{center}
The same happens for a mesh graph. The mesh graph used in the simulations
is a square grid where every intersection represents a node.\\
If the graph is uniformly connected, as in the case of a mesh graph, then the
distribution of single TP fluxes is uniform too.
The majority of single TP fluxes are near to the mean TP
flux, while the highest TP fluxes are shared between different nodes, i.e.
there aren't few unlucky nodes which have to bear a high TP flux.
In the following figure, we can observe the TP flux distribution
of two different QSPN exploration of a mesh graph with $11\times11$ nodes.
In the $x$ axis, an integer point corresponds to a single node of the graph.
Adjacent points are connected by a straight line.\\
\begin{center}
\label{fig:C-dist}
\includegraphics[scale=0.45]{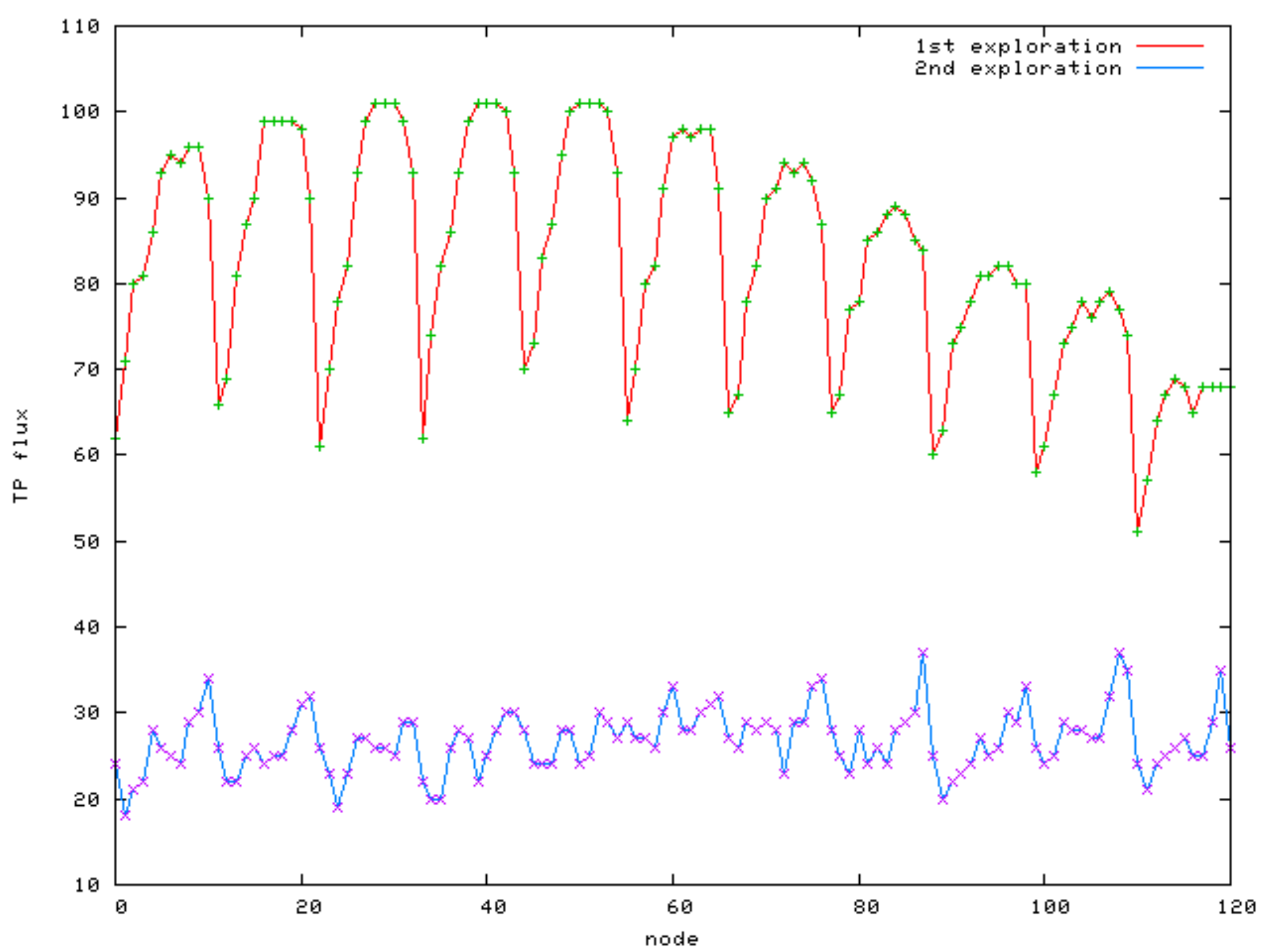}
\end{center}
In the first exploration, plotted in red, the starter node (node 40) and the eight nodes
surrounding him (node 50,51,52,28,29,30,39,41) have the highest TP flux. The TP flux decreases with the
increase of the distance from the starter node. Since the graph is a
$11\times11$ grid, there are $11$ classes where the TP flux falls and for this
reason in the graph there are exactly $11$ slopes. The nodes with lower TP fluxes are
those with the least number of links, i.e. the nodes on the edge of the mesh,
which have only 3 links (and not 4). The four nodes at the vertices of the
grid (node 0,10,110,120), which have just 2
links, register the lowest TP flux.\\
The second exploration, plotted in blue, simulates the scenario where 32 links of the graph
change and consequently the interested nodes send a new CTP to inform the
other nodes and update their maps. As supposed, the mean TP flux (26.80) is heavily
lower than that of the initial exploration (82.90), because the majority of
routes are already known by the nodes. Even the execution time is lower (2.6s
vs 3.5s). Note that this scenario is the most common in real world, because
in general, the nodes won't join or die at once, but progressively.

\section{TODO}
\begin{enumerate}
	\item The $Q^2$ must be implemented in the ntkd daemon.
	\item Improve, test and implement the Caustic Routing:
		\href{http://lab.dyne.org/Ntk\_caustic\_routing}{RFC 0013}
	\item Research a ``mobile QSPN''
\end{enumerate}

\section{ChangeLog}
\begin{itemize}
	\item \verb|April 2007|
		\begin{itemize}
			\item New section: ``Network dynamics'' (\ref{sec:netdyn})
			\item Description of the ETP (sec.  \ref{sec:etp})
			\item Link ID section remove. With the ETP they are no
				more necessary.
			\item More detailed description of the QSPN v1 (sec. \ref{sec:QSPNv1})
			\item Subsection ``QSPN v2 - High levels'' removed. It
				was redundant with the topology
				document\cite{ntktopology}
		\end{itemize}
	\item \verb|October 2006|\\
		Initial release.
\end{itemize}


\newpage

\begin{center}
\verb|^_^|
\end{center}


\begin{thebibliography}{99}
	\bibitem{ntksite} Netsukuku website:
		\href{http://netsukuku.freaknet.org/}{http://netsukuku.freaknet.org/}
	\bibitem{ntktopology} Netsukuku topology document:
		\href{http://netsukuku.freaknet.org/doc/main\_doc/topology.pdf}{topology.pdf}
	\bibitem{DFS} Depth-First Search:
		\href{http://en.wikipedia.org/wiki/Depth-first\_search}{http://en.wikipedia.org/wiki/Depth-first\_search}
	\bibitem{genrouteawk} Generate Routes in Awk:
		\ifpdf \else \\ \fi
		\href{http://cvs.hinezumi.org/viewcvs/netsukuku/proto/doc/qspn/generate\_routes.awk}{generate\_routes.awk}
	\bibitem{simrouteawk} Simplify Routes in Awk:
		\ifpdf \else \\ \fi
		\href{http://cvs.hinezumi.org/viewcvs/netsukuku/proto/doc/qspn/simplify\_routes.awk}{simplify\_routes.awk}
	\bibitem{q2sim} QSPN v2 simulator:
		\ifpdf \else \\ \fi
		\href{http://cvs.hinezumi.org/viewcvs/netsukuku/proto/doc/qspn/q2sim.py}{q2sim.py}
	\bibitem{completegraph} Complete graph:
		\ifpdf \else \\ \fi
		\href{http://mathworld.wolfram.com/CompleteGraph.html}{http://mathworld.wolfram.com/}
	\bibitem{ns2} Network simulator:
		\ifpdf \else \\ \fi
		\href{http://www-mash.cs.berkeley.edu/ns/}{http://www-mash.cs.berkeley.edu/ns/}
	\bibitem{ntkrfc0002} NTK\_RFC 002:
		\href{http://lab.dyne.org/Ntk\_bandwidth\_measurement}{Bandwidth
		measurement}
	\bibitem{aggrsign1}   A Survey of Two Signature Aggregation
		Techniques:
		\ifpdf \else \\ \fi
		\href{http://crypto.stanford.edu/~dabo/abstracts/aggsurvey.html}{http://crypto.stanford.edu/~dabo/abstracts/aggsurvey.html}
	\bibitem{aggrsign2}  Aggregate and Verifiably Encrypted Signatures
		from Bilinear Maps:
		\ifpdf \else \\ \fi
		\href{http://crypto.stanford.edu/~dabo/abstracts/aggreg.html}{http://crypto.stanford.edu/~dabo/abstracts/aggreg.html}

\end{thebibliography}
\end{document}